\documentclass[12pt]{article}

\usepackage{newtxtext,newtxmath}

\usepackage{graphicx}

\usepackage[letterpaper,margin=1in]{geometry}

\renewenvironment{abstract}
	{\quotation}
	{\endquotation}

\date{}

\makeatletter
\renewcommand{\fnum@figure}{\textbf{Figure \thefigure}}
\renewcommand{\fnum@table}{\textbf{Table \thetable}}
\makeatother

\usepackage{scicite}

\usepackage{url}

\def\scititle{
     Scattering near-field optical microscopy at 1-nm resolution using ultralow tip oscillation amplitudes
}
\title{\bfseries \boldmath \scititle}

\author{
	Akitoshi~Shiotari$^{1\ast}$,
	Jun~Nishida$^{2,3}$,
	Adnan~Hammud$^{4}$,
    Fabian~Schulz$^{5}$, \and
    Martin~Wolf$^{1}$,
    Takashi~Kumagai$^{2,3}$,
    Melanie~M{\" u}ller$^{1}$ \and
	\small$^{1}$Department of Physical Chemistry, Fritz-Haber Institute of the Max-Planck Society, \and\small Faradayweg 4--6, 14195 Berlin, Germany. \and
	\small$^{2}$Institute for Molecular Science, National Institutes of Natural Sciences, \and\small 38 Nishigonaka, Myodaichi-cho, 444-8585 Okazaki, Japan.\and
    \small$^{3}$The Graduate University for Advanced Studies, SOKENDAI, Shonan Village, 240-0193 Hayama, Japan.\and
    \small$^{4}$Department of Inorganic Chemistry, Fritz-Haber Institute of the Max-Planck Society, \and\small Faradayweg 4--6, 14195 Berlin, Germany. \and
    \small$^{5}$CIC NanoGUNE, Tolosa Hiribidea 76, 20018 Donostia--San Sebasti{\' a}n, Spain. \and
	\small$^\ast$Corresponding author. Email: shiotari@fhi-berlin.mpg.de 
}

\begin{document} 

\maketitle


\begin{abstract} \bfseries \boldmath

Scattering-type scanning near-field optical microscopy (s-SNOM) allows for the observation of the optical response of material surfaces with a resolution far below the diffraction limit. Based on amplitude-modulation atomic force microscopy (AFM) with typical tapping amplitudes of tens of nanometers, a spatial resolution of 10--100 nm is routinely achieved in s-SNOM. However, optical imaging and spectroscopy of atomic-scale structures remain a substantial challenge. Here, we developed ultralow tip oscillation amplitude s-SNOM (ULA-SNOM), where the ultra-confined field localized at a 1-nm-scale gap between a plasmonic tip and sample is combined with frequency-modulation (non-contact) AFM in a stable cryogenic ultrahigh vacuum environment. Using a silver tip under visible laser illumination with a constant 1-nm amplitude oscillation, we obtain a material-contrast image of silicon islands on a silver surface with 1-nm lateral resolution, which surpasses the conventional limits of s-SNOM. ULA-SNOM paves the way for the acquisition of optical information from atomic-scale structures, such as single photo-active defects and molecules.
\end{abstract}

\section*{Introduction}
\noindent
The combination of optical spectroscopy with scanning tunneling microscopy (STM) enables the optical characterization of material surfaces, nanostructures and molecules as well as their optical control with a resolution far beyond the diffraction limit of light \cite{schultz2020optical,gutzler2021light,muller2023imaging}.
Using plasmomic STM tips, near-field optical techniques in low-temperature (LT) STM \cite{lee2020tip,wang2020fundamental} have been successfully applied to realize optical spectroscopy at the single- or even sub-molecular level, including tip-enhanced Raman spectroscopy \cite{zhang2013chemical,lee2019visualizing}, STM-induced luminescence \cite{qiu2003vibrationally,kuhnke2017atomic}, and tip-enhanced photoluminescence spectroscopy \cite{yang2020sub,imada2021single}. 
Moreover, the plasmonic near-fields in STM junctions allow for controlling single-molecule photoreactions \cite{kazuma2018stm,zhu2022recent,roslawska2024submolecular,PTCDATERS} and visualizing photocurrents through molecular orbitals \cite{imai2022orbital}.
In these works, localized surface plasmon resonances occurring at the nanometer-sized tip apex, so-called nanocavities, are enhanced and confined to a 1-nm$^3$-scale volume by the plasmonic coupling between the tip and sample at a 1-nm scale gap \cite{wang2020fundamental,etchegoin2008perspective,zhang2015optical}.
Additionally, picocavities, formed by the atomistic structure of the tip apex, can provide further spatial confinement of the plasmonic field inside the narrow gap \cite{barbry2015atomistic,benz2016single}.
Such extreme confinement leads to both a localization of the incident light to the atomic scale and a very strong enhancement of optical light emission and scattering from the junction \cite{wang2020fundamental}.
Operation at LT and under ultrahigh vacuum (UHV) conditions also facilitates the stable formation of such 1-nm-scale plasmonic gaps.

In parallel to the aforementioned nanocavity/picocavity-based STM studies, scanning near-field optical microscopy (SNOM) has been established as a standard tool for measuring the local dielectric response of materials. 
The most well-established and sensitive approach is scattering-type SNOM (s-SNOM) \cite{hillenbrand2001pure,chen2019modern}, where amplitude-modulation atomic force microscopy (AM-AFM), also known as tapping-mode AFM, is used to modulate the localized near-field light at the tapping frequency and then detect the demodulated scattering signal at higher harmonics of the tapping frequency using a lock-in amplifier \cite{knoll2000enhanced,raschke2003apertureless}. 
This scheme eliminates contributions from the far-field background and outputs the part of the scattering signal that strongly depends on the tip--sample gap distance. 
This method has enabled the visualization of surface plasmon and phonon polaritons \cite{chen2012optical,fei2012gate,basov2016polaritons,low2017polaritons}, phase transitions \cite{qazilbash2007mott,mcleod2017nanotextured}, and individual biological molecules/complexes \cite{amenabar2013structural,nishida2024sub}, as well as in probing ultrafast dynamics at the nanoscale \cite{jacob2012intersublevel,eisele2014ultrafast,zizlsperger2024situ}. 

Increasing the spatial resolution to the Angstrom scale remains an outstanding challenge in s-SNOM and other scattering-light detection techniques \cite{zenhausern1995scanning,koglin1997material,hillenbrand2002material,huth2012nano,lin2016tip,mastel2018understanding,wang2022high}.
The spatial resolution of conventional s-SNOM is typically limited to tens of nanometers.
While this resolution is sufficient for many applications including the observation of polariton wavelengths longer than $\sim$50 nm \cite{basov2016polaritons}, s-SNOM has not yet accessed more localized structures such as single molecules \cite{betzig1993single} and photo-active point defects \cite{zafar2022recent}.
One of the approaches to achieve high-resolution s-SNOM is to detect higher-harmonics signals, giving rise to weak yet strongly localized response.
Based on this approach, so far 5- or 6-nm resolution was reported as best cases \cite{mastel2018understanding,wang2022high,nishida2024sub}.

As an alternative approach, the use of a sufficiently low amplitude of the cantilever oscillation is expected to be advantageous to sensitively detect light scattering from near-fields localized to the Angstrom scale.
The small tapping essentially enhances the duty cycle for sampling of strongly confined structures, substantially enhancing the sensitivity to localized signals from ultra-narrow tip-sample gaps.
Such an approach was indeed proposed by previous studies on the tapping-amplitude dependence of the s-SNOM images both experimentally \cite{krutokhvostov2011enhanced} and theoretically \cite{esteban2009full,mooshammer2020quantifying}.
However, AM-AFM inherently requires tapping amplitudes greater than $\sim$10 nm to prevent the tapping tip from adhering to the sample surface.
Furthermore, in AM-AFM with low tapping amplitude, the tip motion tends to be anharmonic \cite{mannoni1998measuring,hillenbrand2000higher,bek2005optical}, giving rise to difficulty in interpreting lock-in demodulated signals \cite{esteban2009full}.
These problems can be overcome by frequency-modulation AFM (FM-AFM) \cite{albrecht1991frequency}, also known as non-contact AFM, using a quartz tuning fork (QTF) sensor \cite{giessibl1998high} as a cantilever.
The stiffness of the cantilever and the constant oscillation-amplitude feedback in the FM mode allow for a stable oscillation with a constant, small amplitude ($A \lesssim 1$ nm) \cite{giessibl2019qplus}.
Operation in LT-UHV environments not only leads to high force sensitivity for FM-AFM with high $Q$ values of the oscillation, but also stabilizes an ultranarrow tip--sample gap as used in LT-STM.
Recently, the advantage of combining optical excitation with hybrid STM/FM-AFM systems has been demonstrated in the low-frequency THz regime \cite{siday2024all}.

In this study, we demonstrate FM-AFM-based s-SNOM with an ultralow, 1-nm-scale, cantilever oscillation amplitude, which we refer to as ultralow tip oscillation amplitude s-SNOM (ULA-SNOM).
This enables the generation and detection of extremely localized scattered light from a controlled 1-nm-scale plasmonic gap with unprecedented sensitivity.
Whereas previous works have demonstrated individual aspects such as SNOM with plasmon-resonant tips \cite{fischer1989observation,huth2013resonant,jiang2018near}, at LT \cite{doring2014near,mcleod2017nanotextured,dapolito2022scattering}, with QTF sensors \cite{naber1999dynamic,gucciardi2005interferometric}, or in the FM mode \cite{satoh2017near}, we integrate all those indispensable advances in ULA-SNOM to achieve high-resolution optical imaging.
The combination of s-SNOM and picometer-scale plasmonics paves the way for the future advancement of single-molecule and atomic-scale optical microscopy.

\section*{Results}
\subsection*{ULA-SNOM configuration}

\begin{figure} 
	\centering
	\includegraphics[width=0.8\textwidth]{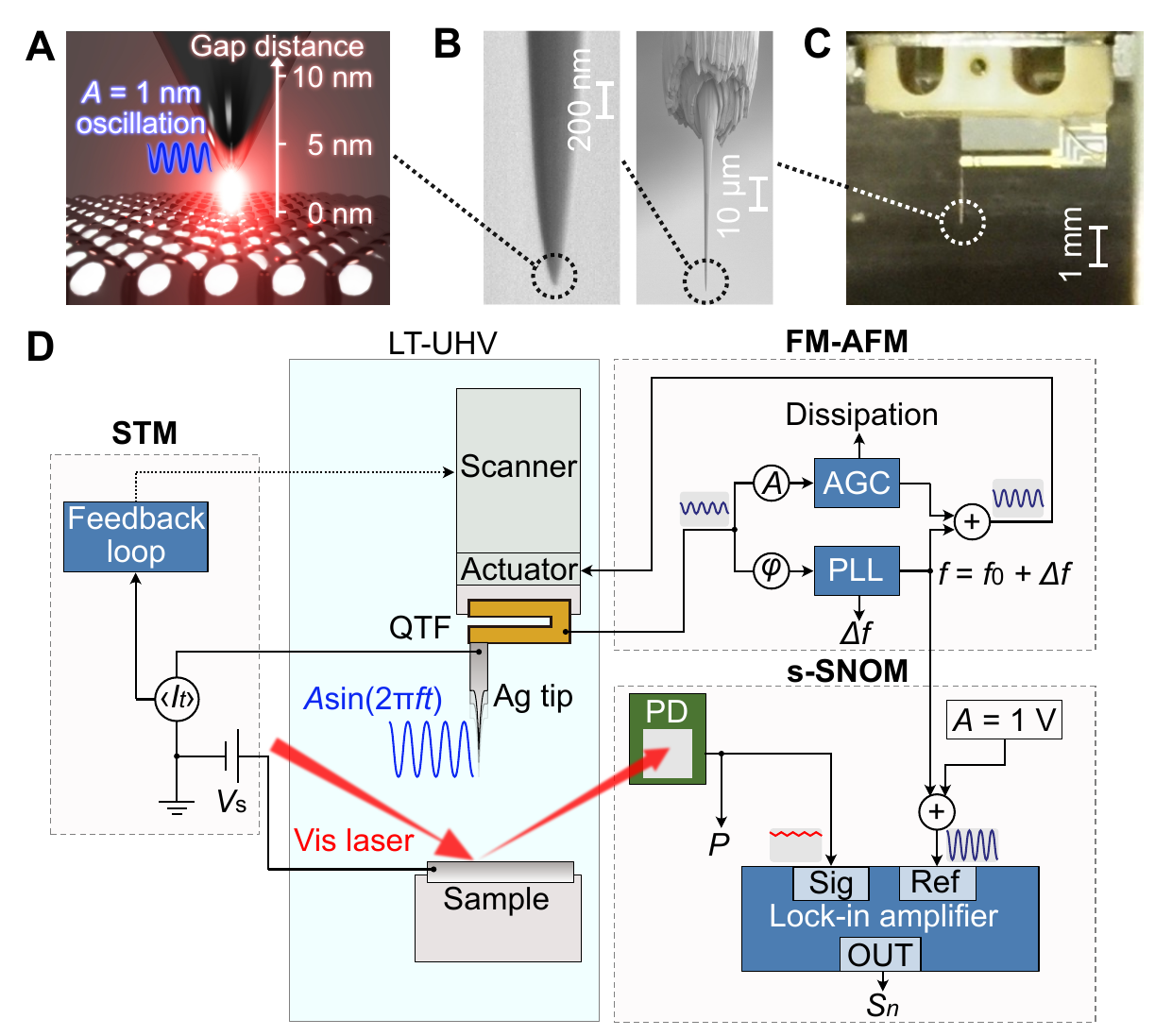} 

	\caption{\textbf{ULA-SNOM setup.}
		(\textbf{A}) Schematic of ULA-SNOM.
        Light scattering from the highly confined picocavity-enhanced near-field can be detected by tip oscillation with an amplitude of 1 nm.
        (\textbf{B}) Scanning electron microscopy images of an Ag tip after the FIB polishing process.
        The left panel shows a magnified image of the tip apex of the image in the right panel. 
        (\textbf{C}) Photo of a QTF sensor with the FIB-polished Ag tip mounted.
        (\textbf{D}) Circuit diagram of ULA-SNOM.
        The STM/FM-AFM unit is located in an UHV chamber at 8 K. 
        A focused 633-nm laser beam illuminates the junction from outside the chamber and scattering light is collected by a PD outside the chamber.
        The PD signal is demodulated by a lock-in amplifier using the cantilever oscillation frequency $f$ as a reference.
    }
	\label{fig1} 
\end{figure}

We performed ULA-SNOM by combining the laser illumination and light detection setup with a commercial LT-UHV STM/FM-AFM setup (Fig.~\ref{fig1}; see also Materials and Methods).
A continuous-wave visible laser beam enters the UHV chamber (wavelength $\lambda$ = 633 nm, incident power $P_\mathrm{inc} = 3$--6 mW, $p$-polarized) and is focused on the tip--sample junction through a lens mounted inside the STM/AFM unit at 8 K (fig.~S1 in the Supplementary Materials).
The light scattered from the junction is collected by a second lens also installed inside the STM/AFM unit and is directed into a photodetector (PD) outside the UHV chamber to measure the power $P$ of the scattered light. 
The apex of an electrochemically etched Ag tip is sharpened and polished by focused ion beam (FIB) milling (Fig.~\ref{fig1}B) to obtain reproducible properties of the plasmonic nanocavity \cite{liu2019resolving,PTCDATERS} and to reduce far-field scattered light from a rough tip shaft.
The tip is mounted on a QTF sensor (Fig.~\ref{fig1}C) which allows for the simultaneous detection of the STM tunneling current $I_t$ and the FM-AFM frequency shift $\Delta f$.
When the cantilever is oscillated, the measured tunneling current is the time average over the oscillation cycle, denoted as $\langle I_\mathrm{t} \rangle$.
The current signal is used for feedback control of the tip height $z$ (Fig.~\ref{fig1}D) as general in QTF-based STM/FM-AFM \cite{giessibl2019qplus}.
The cantilever is resonantly oscillated with a frequency $f = f_0 + \Delta f$, which is fed back by an automatic gain controller (AGC) to keep the oscillation amplitude $A$ constant and by a phase-locked loop (PLL) to obtain $\Delta f$ induced by tip--sample interactions \cite{albrecht1991frequency}. 
We use a sine-wave output $\sin(2 \pi ft)$ from the oscillation controller as the reference signal for the lock-in detection in real time (Fig.~\ref{fig1}D), obtaining $n$-th harmonic signals $S_n$ from the PD output.
We verified the pure harmonic cantilever oscillation even at a tunneling regime (fig.~S2), eliminating the possibility of artifacts in the $S_n$ signals due to higher harmonic components of the cantilever oscillation \cite{esteban2009full}.

\subsection*{Simultaneously recorded STM, FM-AFM, and s-SNOM signals}

\begin{figure} 
	\centering
	\includegraphics[width=0.65\textwidth]{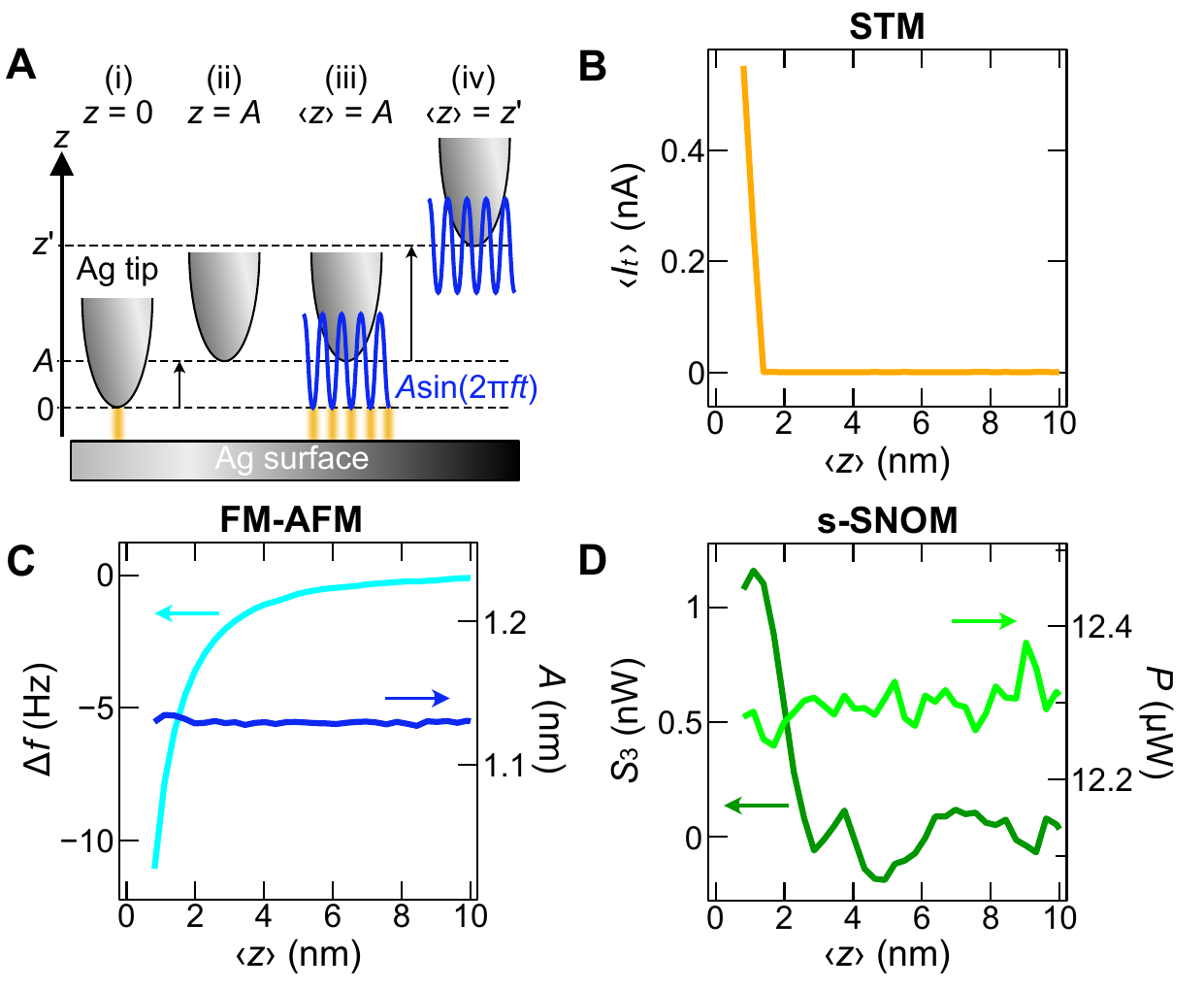} 

	\caption{\textbf{Approach curves of STM, FM-AFM, and s-SNOM.}
	(\textbf{A}) Schematic side-view of the way to obtain an approach curve with an oscillation amplitude $A$.
    The origin of $z$ is defined as the tip height determined by the STM set-point (sample bias $V_\mathrm{s}$ = 30 mV and tunneling current $I_\mathrm{t}$ = 0.10 nA) without cantilever oscillation, as depicted in (i).
    After the tip retraction by $A$ (ii), the oscillation is started (iii).
    At the point, the tip height is described as $z(t) = A [1+\sin(2 \pi ft)]$, where $\langle z(t) \rangle = A$ and $\mathrm{min}[z(t)] = 0$.
    Then the tip is moved vertically to a given point $\langle z \rangle = z'$ (iv) to record an approach curve.
    The orange lines between the tip and sample depict tunneling current.
    (\textbf{B} to \textbf{D}) Approach curves of STM ($\langle I_\mathrm{t} \rangle$), FM-AFM ($A$ and $\Delta f$), and s-SNOM ($P$ and $S_3$), respectively, recorded over an Ag(111) terrace ($P_\mathrm{inc} = 3$ mW, $A = 1.13$ nm).
    }
	\label{fig2} 
\end{figure}

The localization of the near-field light inside the junction can be characterized by measuring tip-approach curves, which represents the tip-height dependence of the scattered signal at a given oscillation amplitude $A$.
Figure~\ref{fig2}A shows the measurement procedure.
First, the Ag tip is placed over an atomically flat Ag(111) surface without cantilever oscillation and with the STM feedback closed (defined as $z = 0$; (i) in Fig.~\ref{fig2}A). 
Next we open the feedback loop and retract the tip from $z=0$ to $z=A$ (ii). 
We then start the sinusoidal cantilever oscillation with an amplitude of $A$ such that the tip oscillated around the center position $z=A$ (iii).
Therefore, the lowest tip height during the oscillation is equal to the set-point distance, i.e., $\mathrm{min}[z(t)] = 0$, ensuring that the tip does not crash into the surface when varying $A$.
The instantaneous tip height during oscillation is expressed as $z(t) = \langle z \rangle + A \sin(2\pi ft)$, where $\langle z \rangle$ is the time-averaged tip height equal to the center of oscillation.
We then sweep the tip height to acquire approach curves in the range between $\langle z \rangle = A$ and a given maximum height $z'$ ((iv) in Fig.~\ref{fig2}A). 
The approach curve of the s-SNOM signal $S_n(\langle z \rangle)$ is obtained along with the STM and FM-AFM signals.

Figures~\ref{fig2}B to \ref{fig2}D show the STM signal ($\langle I_\mathrm{t} \rangle$), FM-AFM signals ($A$ and $\Delta f$), and the scattered laser power $P$ together with the third-harmonics s-SNOM signal $S_3$, respectively, simultaneously recorded over a Ag terrace with a set-point oscillation amplitude of 1.13 nm and during 633-nm laser illumination.
Because a typical metal-tip--metal-sample gap distance during STM feedback is approximately 0.5 nm \cite{zhang2011quantum}, we estimate the gap at $z = 1$ nm to be $\sim$1.5 nm.
We obtained the curves at $\langle z \rangle = 1$ to 150 nm.
Only plots at $\langle z \rangle \leq 10$ nm are displayed in Fig.~\ref{fig2} because at larger tip heights, no signal exceeding the noise floor was observed in $\langle I_\mathrm{t} \rangle$, $\Delta f$, or $S_3$ (fig.~S3).
The time-averaged current $\langle I_\mathrm{t} \rangle$ was detectable only at very close tip heights and shows a steep rise at $\langle z \rangle < 1.5$ nm as expected from the exponential $z$ dependence of tunneling current (Fig.~\ref{fig2}B). 
The FM-AFM system always keeps $A$ constant (Fig.~\ref{fig2}C, right axis) whereas $\Delta f$ decreases monotonically as $\langle z \rangle$ decreases (left axis) due to attractive forces between the Ag tip and Ag sample.
These tip-height dependences are consistent with standard STM/FM-AFM operation \cite{ternes2011interplay}, indicating that the laser illumination does not interfere the STM/FM-AFM performance.

As shown in the PD signal $P$ (Fig.~\ref{fig2}D, right axis), the near-field contribution to the total scattered power is faint and obscured by the noise of the PD channel, implying that $P$ is dominated by background scattering with a constant intensity.
However, lock-in detection at the third harmonic $S_3$ of the oscillation frequency (Fig.~\ref{fig2}D, left axis) allows for the extraction of the near-field enhancement occurring only at very small gap distances.
The steep increase of $S_3$ at $\langle z \rangle < 3$ nm is consistent with the spatial confinement of electromagnetic fields in the tip--sample gaps \cite{wang2020fundamental,etchegoin2008perspective,barbry2015atomistic}.
Notably, the STM, FM-AFM, and s-SNOM channels exhibit different tip-height thresholds for signals exceeding their noise floors ($\langle z \rangle \approx 1.5$ nm for $\langle I_\mathrm{t} \rangle$, $\sim$7 nm for $\Delta f$, and $\sim$3 nm for $S_3$) as well as different steepness of the slopes, showing that their signals are independent and of different origins. 

\subsection*{Optimal oscillation amplitude for ULA-SNOM}

\begin{figure} 
	\centering
	\includegraphics[width=0.6\textwidth]{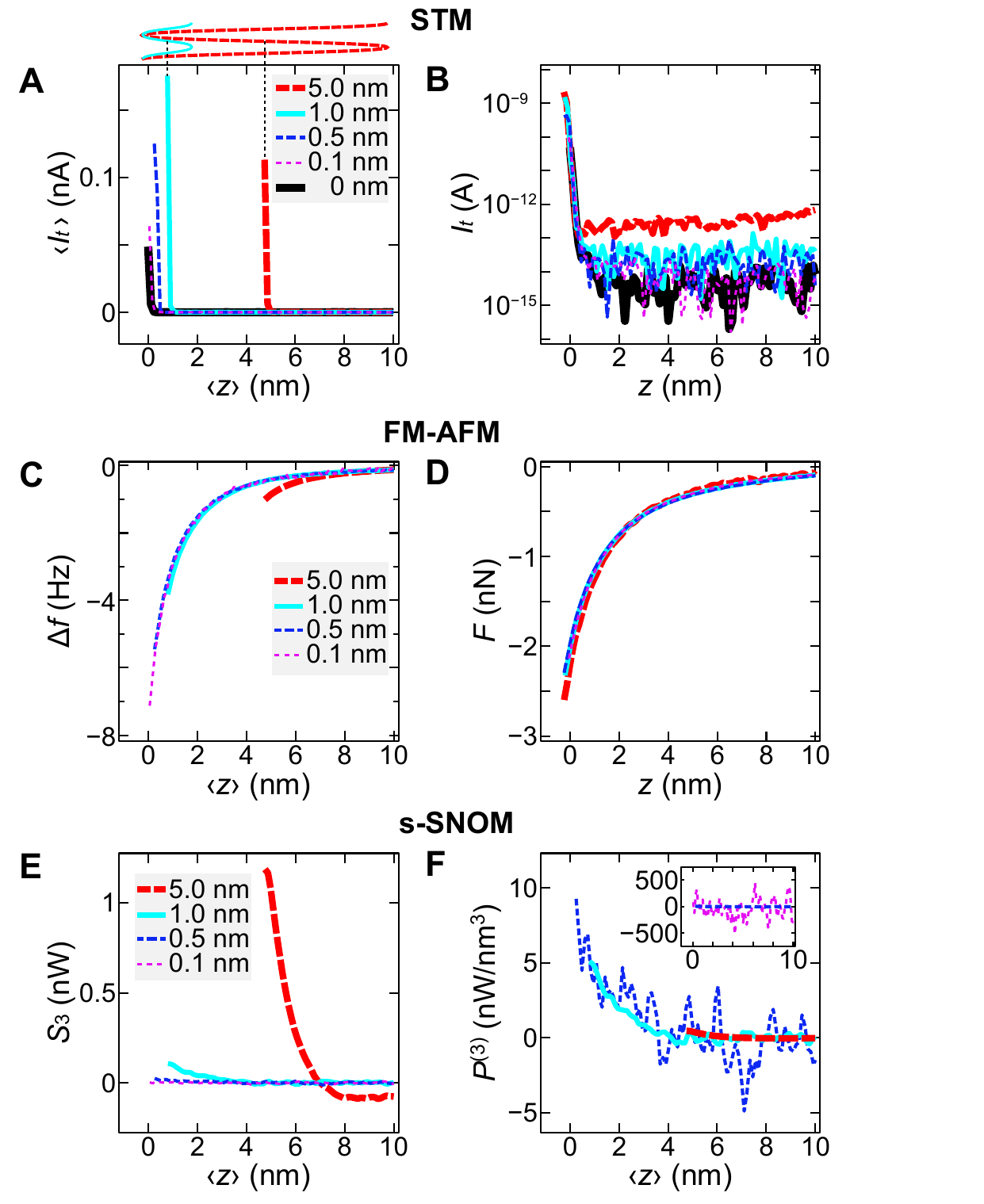} 

	\caption{\textbf{Amplitude-dependent approach curves and their converted/normalized curves.}
	(\textbf{A}) Approach curves of $\langle I_\mathrm{t} \rangle$ recorded over an Ag(111) terrace for different oscillation amplitudes $A$ ($P_\mathrm{inc} = 6$ mW).
    The sine waves on top of the graph illustrate the tip trajectories during the cantilever oscillation with $A = 5.0$ nm (red dotted curve) and 1.0 nm (cyan solid) at the minimum $\langle z \rangle$ points in the plots.
    (\textbf{B}) $I_\mathrm{t}(z)$ curves converted from (A).
    (\textbf{C}) Approach curves of $\Delta f$ simultaneously recorded with (A).
    (\textbf{D}) $F(z)$ curves converted from (C).
    (\textbf{E}) Approach curves for $S_3$ simultaneously recorded with (A).
    (\textbf{F}) Normalized s-SNOM curves calculated from (E).
    The normalized curve with the lowest amplitude ($A = 0.1$ nm) is not shown in the main graph of (F) but in the inset (magenta dotted curve) because of the large noise.
    As a reference, the curve with $A = 0.5$ nm (blue dashed curves) is also shown in the inset. 
    With any $A$, the origin of $z$ is defined as the tip height determined by the STM set-point at $V_\mathrm{s}$ = 30 mV, $I_\mathrm{t}$ = 0.10 nA without oscillation (see Fig.~\ref{fig2}A). 
    }
	\label{fig3} 
\end{figure}

As mentioned above, a small value of $A$ is advantageous for detecting highly localized signals.
However, if $A$ is too small, this will reduce the signal-to-noise ratio in the lock-in detection.
It is therefore important to examine the optimal oscillation amplitude to obtain the best localization at a reasonably high signal level. 
For this purpose, we measure approach curves with various values of $A$.
Figures~\ref{fig3}A, C, and E show the approach curves of $\langle I_\mathrm{t} \rangle$, $\Delta f$, and $S_3$, respectively, recorded simultaneously over the Ag(111) surface for $A$ between 0.1 and 5.0 nm. 
As a reference, $I_\mathrm{t}(z)$ curve without cantilever oscillation is also shown (Fig.~\ref{fig3}A, bold black curve).
At the largest amplitude $A = 5.0$ nm, all signals exceed their noise floors at much larger tip heights (for example, $\langle z \rangle = 5$ nm for $\langle I_\mathrm{t} \rangle$) compared to the plots at lower $A$ values (see also Fig.~\ref{fig2} at $A = 1.13$ nm).
This behavior can be explained by tip trajectory during oscillation, as illustrated by the sine waves depicted on top of Fig.~\ref{fig3}A. 
At the closest $\langle z \rangle$, which depends on $A$ as described above (Fig.~\ref{fig2}A), the tip temporarily approaches the STM set-point distance during each oscillation cycle, giving rise to signals in $\langle I_\mathrm{t} \rangle$, $\Delta f$, and $S_3$.
The approach curves of $\langle I_\mathrm{t} \rangle$ and $\Delta f$ can be converted into the instantaneous tunneling current $I_\mathrm{t}$ (Fig.~\ref{fig3}B) and vertical force $F$ (Fig.~\ref{fig3}D) at the bottom of the oscillation, i.e., $z = \mathrm{min}[z(t)]$, using the Sader--Sugimoto \cite{sader2010accurate} and Sader--Jarvis \cite{sader2004accurate} formulae, respectively.
The converted curves are consistent at any $A$, corroborating the stable, harmonic motion of the cantilever oscillation, which is critical to extract reliable s-SNOM signals upon demodulation \cite{esteban2009full}.
Notably, the conversions allow for the evaluation of the tunneling conductivity and interatomic force in narrow tip--sample gaps, e.g., at $z = 0$, using any $A$.

For s-SNOM, large $A$ limits the information at narrow tip--sample gaps, unless sufficiently high harmonic channels are used \cite{wang2022high,nishida2024sub}.
Here, we discuss the accessibility of s-SNOM signals in a small $z$ range by normalizing the $A$-dependent intensity of the lock-in signals. 
The scattered light intensity $P$ which is modulated by the cantilever oscillation can be expressed by a Taylor series as
\begin{eqnarray}
    P[z(t)] &=& P[ \langle z \rangle + A \sin(2\pi ft) ]  \nonumber \\
         &=& \sum^{\infty}_{m=0} \frac{(-A)^m}{m!} P^{(m)}(\langle z \rangle) \sin^m (2\pi ft),
          \label{eqP}
\end{eqnarray}
\noindent
where $P^{(m)}(\langle z \rangle) \equiv (-1)^m \left.  \frac{\mathrm{d}^mP}{\mathrm{d}z^m} \right|_{z= \langle z \rangle}$.
The lock-in signal corresponds to the time-averaged value of the input signal $P$ multiplied with a sine-wave reference signal with a frequency of $nf$ for the $n$-th harmonics, i.e.,
\begin{equation}
    S_n(\langle z \rangle) = \left\langle P[z(t)] \sin(2\pi nft + \phi_n)  \right\rangle,
    \label{eqP_timeavg}
\end{equation}
\noindent
where $\phi_n$ denotes the phase difference between the input and reference signals (see Supplementary Text 1).
From Eqs.~\ref{eqP} and \ref{eqP_timeavg}, $S_n$ is solved as 
\begin{equation}
    S_n(\langle z \rangle)
        = \frac{1}{n! 2^n} A^n P^{(n)}(\langle z \rangle) 
        + \sum^{\infty}_{i=1} c_{n+2i}A^{n+2i} P^{(n+2i)}(\langle z \rangle),
        \label{eqSn} 
\end{equation}
\noindent
where $c_{n+2i}$ denotes the coefficient for the $(n+2i)$-th derivative component, which is much smaller than $\frac{1}{n! 2^n}$ (see Supplementary Text 1).
In the case of FM-AFM, $A$ is constant at any $z$ and a sufficiently low $A$ eliminates the contribution of higher order terms.
Therefore, the $S_3$ approach curve (Fig.~\ref{fig3}E) can be normalized as $P^{(3)}(\langle z \rangle) = 48 A^{-3} S_3(\langle z \rangle)$ (Fig.~\ref{fig3}F; see also fig.~S4 for other $n$-th harmonics curves and their normalization).

After normalization, all approach curves recorded at different $A$ yield the same exponential $\langle z \rangle$ dependence of the near-field signal (Fig.~\ref{fig3}F), however with different noise levels and minimum distances $\langle z \rangle$.
The largest amplitude ($A = 5.0$ nm) provides high signal-to-noise ratio, but it misses the information below $\langle z \rangle \approx 5$ nm and does not allow to extract the picocavity-enhanced near-field signal. 
Furthermore, at such a large $A$, the contribution of the higher order derivative components (see the second term on the right side of Eq.~\ref{eqSn}) to $S_n$ is no longer negligible, modifying the curve appearance from the original $P^{(n)}(\langle z \rangle)$ curve shape (see Supplementary Text 1).
For these reasons, too large values of $A$ are not suitable for the relatively low harmonics detection, such as $n = 1$--4.
In contrast, at the smallest amplitude $A = 0.1$ nm, no s-SNOM signal $S_3$ was measurable above the noise floor (inset in Fig.~\ref{fig3}F). 
Its normalized curve has an approximately 50 times higher noise intensity than the signals detected at higher $A$.  
Therefore, we conclude that amplitudes of 0.5--1 nm are optimal for the ULA-SNOM experimental setup used. 
Note that this value depends on the intensity of light scattering from the near-field and the signal sensitivity of the light detection setup.
Higher collection efficiency of the scattered light is expected to allow for s-SNOM signal detection with smaller $A$.

\subsection*{s-SNOM imaging of Si monolayer islands on Ag(111)}

To demonstrate the lateral resolution and optical contrast of ULA-SNOM, we use an Ag(111) surface partially covered by ultrathin Si islands (Fig.~\ref{fig4}A).
According to the STM appearance, the islands, which partially cover the terraces and step edges (see fig.~S5A for the overview STM image), are ascribed to amorphous Si films \cite{solonenko2017comprehensive}.
Note that the islands appear darker in the image than the Ag terrace despite being located on the terrace.
This contrast originates from the lower local density of states of Si compare to Ag, which competes with the topographic height difference.
The coexistence of the Si monolayer islands and bare Ag surfaces minimizes the topographic height difference between Si and Ag (Fig.~\ref{fig4}G), while retaining the plasmon enhancement effect between the Ag tip and Ag substrate \cite{schultz2022characterizations}. 

We record simultaneous STM, FM-AFM and s-SNOM images of the sample (Figs.~\ref{fig4}A to \ref{fig4}F).
Both the QTF-based FM-AFM and the high harmonic lock-in detection require rather long acquisition times ($\sim$30 times slower than the scan speed in standard STM imaging mode), rendering imaging over larger areas at constant tip height challenging.
Therefore, the tip height was controlled by the STM feedback loop to avoid thermal drift during the slow scan.
As schematically shown in Fig.~\ref{fig4}G, the scanned area has two Si islands partially covering a Ag terrace.
Under the STM feedback, the Ag-tip--Ag-surface gaps are slightly narrower by $\sim$50 pm than the Ag--Si gaps (the topmost line profile in Fig.~\ref{fig4}I and the orange dotted curve in Fig.~\ref{fig4}G). 
The FM-AFM $\Delta f$ map (Fig.~\ref{fig4}B) shows several dark spots inside the Si islands, presumably due to Si atoms/clusters in the Si monolayer attractively interacting with the Ag tip \cite{majzik2013combined}.
On the other locations in the Si island, the $\Delta f$ value is similar to that over Ag, and forces applied over Ag and Si are also comparable (see fig.~S6 for the $\Delta f(\langle z \rangle)$ and $F(z)$ curves).
The different appearance in each map suggests that simultaneous STM/FM-AFM/s-SNOM mapping provides complementary information on the scanned area due to the different signal origin of each image.

Figures~\ref{fig4}C to \ref{fig4}F show s-SNOM images for different harmonics $S_n$. 
Interestingly, the image appearance changes with $n$.
The images of $S_1$ and $S_2$ (Figs.~\ref{fig4}E and \ref{fig4}F) are not sensitive to the presence of the Si islands.
The middle plot of Fig.~\ref{fig4}I shows the line profile of the $S_1$ map over a boundary between Ag and Si (dotted line in Fig.~\ref{fig4}C), where the boundary gives no signal change.
The Si island on the left side in the maps (Fig.~\ref{fig4}C and D) is imaged slightly brighter than the bare Ag, which is presumably attributed to the topographic artifact; the tip height over Si is slightly lower than Ag (Fig.~\ref{fig4}G), providing the faint signal difference.
The topographic effect on $S_1$ and $S_2$ was confirmed by mapping across an Ag step (see Supplementary Text 2).
On the contrary, at higher harmonics $S_3$ (Fig.~\ref{fig4}E) and $S_4$ (Fig,~\ref{fig4}F), the image contrast changes as the Si islands now exhibit darker than the Ag terraces.
This contrast is opposite of the topographic artifact effect, strongly indicating that the images of $S_3$ and $S_4$ are sensitive to changes in the local dielectric environment caused by the Si islands.

The $n$ dependence of the s-SNOM appearance agrees with the common understanding that the localized near-field signal is predominantly detected at high $n$ \cite{knoll2000enhanced,hillenbrand2002material,wang2022high,nishida2024sub}.
The lower harmonics are not sensitive to the ultra-confined near-field in the 1-nm-scale gap. 
Hence, whereas the image appearances of $S_1$ and $S_2$ are dominated by topographic artifacts, the images of $S_3$ and $S_4$ show true optical, i.e. dielectric, material contrast. 
This is further corroborated by the observation that an atomic-scale structural change in the tip apex modifies the magnitude of the material contrast (see Supplementary Text 2 and fig.~S5).

Figure~\ref{fig4}H shows $S_4$ approach curves recorded over a Ag terrace and a Si island on it, also indicating that the $S_4$ signal over Si is smaller than that over Ag at small $\langle z \rangle$.
This trend is consistent with previous calculations \cite{hillenbrand2002material,raschke2003apertureless} reporting that at 633 nm, Ag gives a larger scattering intensity in $S_n$ than Si due to the difference in the real part of the dielectric constant ($-18+i0.5$ for Ag versus $15+i0.2$ for Si). 
Considering that the image contrast originates only from a single atomic layer of Si, the contrast between the two surface regions is quite striking.

We note that our implementation of ULA-SNOM is based on a self-homodyne scheme (see fig.~S1), which arises from the interference between tip-localized scattering and background scattering \cite{chen2019modern}.
In our s-SNOM imaging, nevertheless, the length scales of the cantilever oscillation amplitude, tip--sample gap distance, and scanned size are much smaller than the wavelength of the light ($A < \lambda/600$), where the amplitude and phase of the background field should be constant.
This in part is facilitated by the FIB-polished tip and the atomically flat single-crystalline surface under UHV conditions, which substantially reduce uncontrolled scattering.
Furthermore, the aforementioned results of the $n$-dependent s-SNOM images (Figs.~\ref{fig4}C to \ref{fig4}F) and the consistency of the Ag/Si contrast with the numerical prediction \cite{hillenbrand2002material,raschke2003apertureless} 
support that the background effect alone cannot account for the observed contrast. 
ULA-SNOM is, in principle, compatible with interferometric detection methods such as pseudo-heterodyne, which have been established to provide amplitude- and phase-resolved responses in a fully background-free manner \cite{ocelic2006pseudoheterodyne,chen2019modern}. 
Combining ULA-SNOM with such interferometric techniques offers the potential for a full characterization of the dielectric function at the atomic scale. 

Finally, we estimate the lateral resolution of s-SNOM and compare it to that of STM by taking line profiles along the images across a material boundary. 
The bottommost plot in Fig.~\ref{fig4}I shows the line profile of the $S_4$ signal recorded simultaneously with the STM topography (topmost plot).
The faint oscillation spanning over a few nanometers in the STM line profile presumably originates from Friedel oscillations on Ag(111) \cite{hasegawa1993direct}. 
Both profiles exhibit step shapes at the boundary between Ag and Si, in contrast to the profile of $S_1$ without material contrast (middle plot in Figure~\ref{fig4}I). 
To quantitatively evaluate the lateral resolution, we conducted the peak fitting analysis of the line profiles, as in previous studies \cite{hillenbrand2002material,lin2016tip,mastel2018understanding}.
We use an error function,
$y(x) = y_0 + c \, \mathrm{erf}\left( \frac{x-x_0}{\sqrt{2} \sigma} \right)$, where $y(x)$ is the signal profile, $y_0$ and $c$ are the offset and step-height coefficient, respectively, $x_0$ is the boundary position, and $\sigma$ denotes the full width at half maximum.
The fitting (solid curves in Fig.~\ref{fig4}I) results in $\sigma = 0.39 \pm 0.26$ for STM and $1.06 \pm 0.13$ nm for s-SNOM $S_4$.
The observed difference in the spatial resolution between STM and s-SNOM provides further evidence that the contrast in the higher-harmonics s-SNOM images is not caused by topographic tip-height changes or tip motion \cite{hecht1997facts}, but indeed reveals true optical contrast.  
The ability to resolve optical contrast with a spatial resolution as small as 1 nm in elastic light scattering provides an approach for optical surface analysis at the nearly atomic scale.

\section*{Discussion}

We demonstrated the successful implementation and development of ULA-SNOM employing light scattering from a stable plasmonic tip--sample nanojunction at a cryogenic temperature. 
We conducted FIB-polishing of a plasmonic Ag tip mounted to the tuning fork sensor of FM-AFM and show that the highly confined near-field in the 1-nm-scale gap can be detected by demodulation and lock-in detection of the scattering signal at higher harmonics of the oscillation frequency of the QTF. 
We image the optical contrast between a bare Ag(111) surface and monoatomic Si islands on the surface with a lateral resolution of 1 nm using the fourth harmonics of the cantilever oscillation frequency. 
ULA-SNOM will be of interest for the optical characterization of a wide range of conducting and insulating nanomaterials exhibiting optical and dielectric heterogeneity at the atomic scale.
The integration of s-SNOM in the LT-UHV STM/FM-AFM setup enables the simultaneous detection of multiple independent observables including electric conductivity (STM), interatomic attractive/repulsive forces (FM-AFM), and dielectric constant (s-SNOM), which promises new insights into, e.g., the photophysics of single defects and molecules and the optical properties of atomically sharp interfaces. 



\subsection*{Materials and Methods}

\subsection*{ULA-SNOM setup}
ULA-SNOM was performed with an LT-UHV STM/FM-AFM machine (CreaTec Fischer \& Co.~GmbH; base pressure $< 5 \times 10^{-11}$, a sample temperature of 8 K). 
The optical configuration on a self-homodyne scheme is depicted in fig.~S1A.
The output voltage of the Si-biased PD was converted into the light power $P$ by the calibration with a Si-photodiode laser power meter.
We confirmed that comparable results were obtained with another optical configuration of a back-scattering geometry (figs.~S1B to S1E).
As a QTF sensor for the FM-AFM operation, we used a qPlus sensor \cite{giessibl1998high} on which a Ag tip was mounted (CreaTec Fischer \& Co.~GmbH; spring constant of 1800 N/m, Q value of $\sim$10$^4$, sample-free resonance frequency $f_0$ of 18.8 kHz; Fig.~\ref{fig1}C) and an oscillation controller (Nanonis OC4, SPECS Surface Nano Analysis GmbH).
The sine-wave output with a frequency $f$ from the controller was connected to the reference-signal input of a lock-in amplifier (HF2LI, Zurich Instruments), demodulating the PD signals (Fig.~\ref{fig1}D).
The quadrature (out-of-phase) component of each harmonics was always zero (see Supplementary Text 1), while each in-phase components were monitored as an s-SNOM channel.

\subsection*{Tip fabrication}

We sharpened and polished the apex of the electrochemically etched Ag tip mounted on the sensor (Fig.~\ref{fig1}B) by Ga$^+$ FIB milling.
During the FIB process, the tunneling-current electrode of the sensor was grounded to prevent charging of the tip.
Notably, while the FIB fabrication for STM tips have been reported \cite{bockmann2019near,liu2019resolving,PTCDATERS}, this is the first report for the FIB process of a tip mounted on a QTF sensor.
After the sensor with the FIB-polished tip was introduced into the UHV chamber, the tip apex was further adjusted by mild poking into a clean Ag surface by the STM-based tip-height control to get a strong plasmon resonance \cite{PTCDATERS}.

\subsection*{Sample fabrication}

\begin{figure} 
	\centering
	\includegraphics[width=0.74\textwidth]{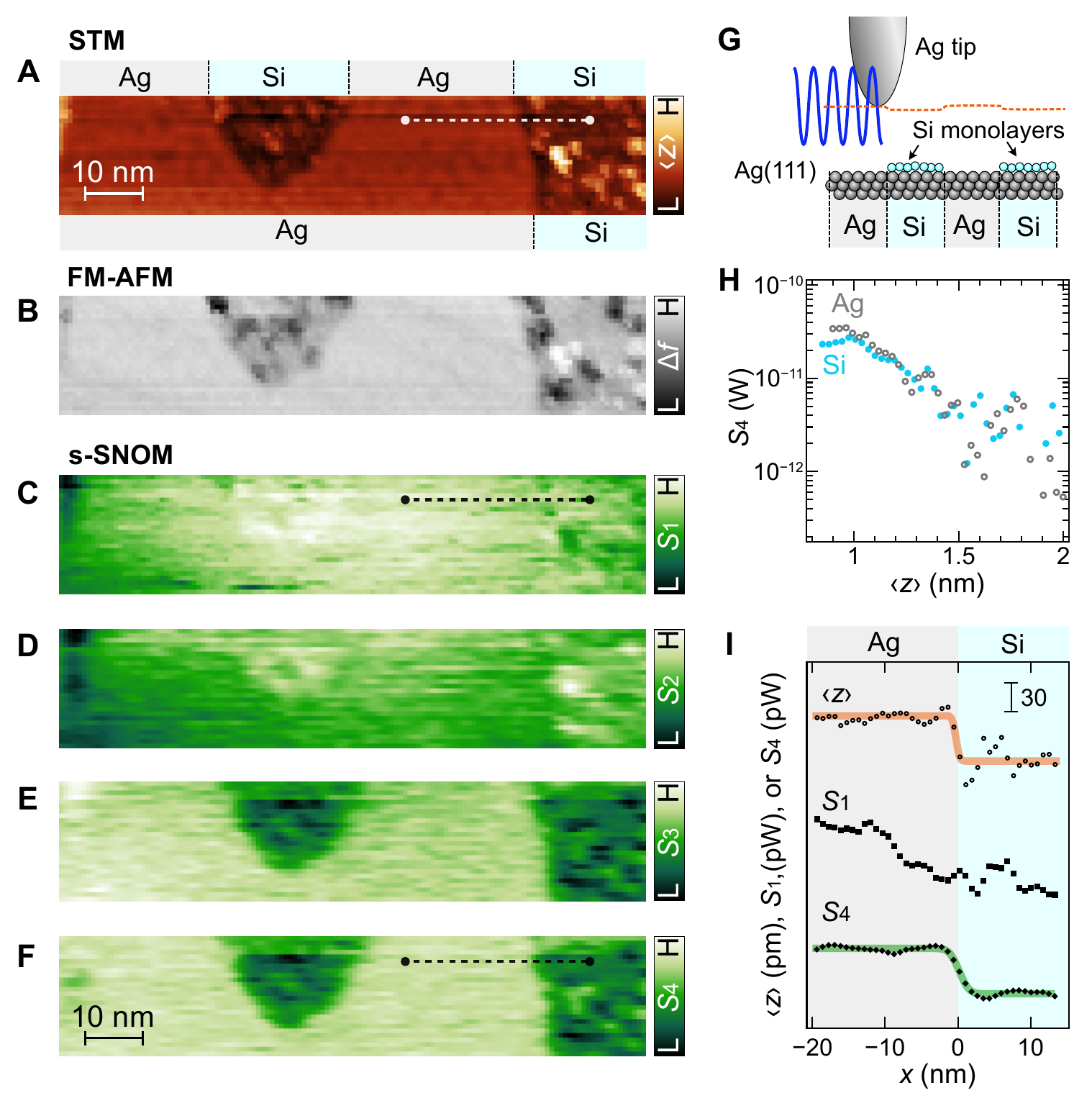} 

	\caption{\textbf{Simultaneously acquired STM, FM-AFM, and s-SNOM images.}
    (\textbf{A}) STM topography, (\textbf{B}) FM-AFM $\Delta f$ map, and (\textbf{C} to \textbf{F}) s-SNOM $S_1$ to $S_4$ maps simultaneously obtained (STM set-point: $V_\mathrm{s}$ = 30 mV, $\langle I_\mathrm{t} \rangle$ = 0.10 nA, $A$ = 1.0 nm; $P_\mathrm{inc} = 6$ mW).
    The highest (H) and lowest (L) values of each color bars are as follows: (H, L) = (0.32, $-$0.15) nm (A), ($-$0.31, $-1.6$) Hz (B), (21, 19) nW (C), (235, 19) pW (D), (151, 31) pW (E), and (51, $-$49) pW (F).
    (\textbf{G)} Side-view scheme of the atomic structures of the sampling area.
    Labels ``Ag'' and ``Si'' in A and G represent the bare Ag terrace and Si island on the terrace, respectively.
    (\textbf{H}) Approach curves of $S_4$ recorded over an Ag terrace (gray empty bullets) and an Si island on it (cyan filled bullets), recorded with another Ag tip (fig.~S5H) than that for the maps in A--F. 
    The origin of $\langle z \rangle$ for both plots is defined by the STM set-point over the Ag terrace (set-point: $V_\mathrm{s}$ = 30 mV and $I_\mathrm{t}$ = 0.10 nA without oscillation). 
    (\textbf{I}) Line profiles of the STM topographic height $\langle z \rangle$ (top), $S_1$ (middle), and $S_4$ (bottom) across an Ag--Si boundary on an identical terrace.
    The dotted lines in A, C, and F indicate the sampling position.
    The solid curves are the fitting curves for $\langle z \rangle$ (top) and $S_4$ (bottom) with error functions.
    }
	\label{fig4} 
\end{figure}

We used a single-crystalline Ag(111) surface (MaTeck GmbH) cleaned by multiple cycles of Ar$^+$ sputtering and annealing.
For the Si/Ag sample preparation (Fig.~\ref{fig4}A), we used a home-made Si evaporator placed $\sim$30 cm away from the cleaning stage for the Ag(111) surface in the UHV chamber.
The evaporator has a direct-current-heated Si(111) plate (Siegert Wafer GmbH) flash-annealed at 1200 $^\circ$C.
During the Si evaporation, the cleaned Ag(111) sample was heated at 227$^\circ$C to 230$^\circ$C and faced the evaporator.


\clearpage 

%

%
%
%
%
%
%


\section*{Acknowledgments}
\paragraph*{Funding:}
F.S.~acknowledges financial support from grant PID2022-140845OBC61 and the Ram{\' o}n y Cajal fellowship RYC2021-034304-I.
T.K.~acknowledges the support of JST FOREST Grant No.~JPMJFR201J.
\paragraph*{Author contributions:}
A.S., M.W., T.K., and M.M.~directed the project.
A.S., J.N., and T.K.~designed and constructed the experimental setup.
A.H.~conducted the FIB-polishing process of the Ag tip mounted on the QTF sensor.
A.S.~performed the experiments and analyzed the data.
A.S., F.S., M.W., and M.M.~discussed the data interpretation.
A.S.~wrote the manuscript and all authors participated in revisions of the manuscript.
\paragraph*{Competing interests:}
The authors declare they have no competing interest.
\paragraph*{Data and materials availability:}
All data needed to evaluate the conclusions in the paper are present in the paper and/or the Supplementary Materials.


\subsection*{Supplementary materials}
Supplementary Text\\
Figs.~S1 to S6 \\
References \cite{hembacher2004force,giessibl2006higher}


\newpage


\renewcommand{\thefigure}{S\arabic{figure}}
\renewcommand{\thetable}{S\arabic{table}}
\renewcommand{\theequation}{S\arabic{equation}}
\renewcommand{\thepage}{S\arabic{page}}
\setcounter{figure}{0}
\setcounter{table}{0}
\setcounter{equation}{0}
\setcounter{page}{1} 


\begin{center}
\section*{Supplementary Materials for\\ \scititle}

    Akitoshi~Shiotari$^{\ast}$, 
	Jun~Nishida, 
	Adnan~Hammud, 
    Fabian~Schulz, 
    Martin~Wolf, 
    Takashi~Kumagai, 
    Melanie~M{\" u}ller \and 
    \\
	\small$^\ast$Corresponding author. Email: shiotari@fhi-berlin.mpg.de 
\end{center}

\subsubsection*{This PDF file includes:}
Supplementary Text\\
Figures S1 to S6 \\
References \cite{hembacher2004force,giessibl2006higher}

\newpage


\subsection*{Supplementary Text}

\subsubsection*{Supplementary Text 1: Normalization of s-SNOM approach curves}

Here, we show the detail of the Taylor series for the scattered light intensity $P$ modulated by the cantilever oscillation with an amplitude of $A$ and a frequency of $f$ (Eq.~1 in the main text);
\begin{eqnarray}
    P[z(t)] &=& P[ \langle z \rangle + A \sin\theta ]  \nonumber \\
        &=&  P(\langle z \rangle)  \nonumber \\
        &&+\ A \left.  \frac{\mathrm{d}P}{\mathrm{d}z} \right|_{z= \langle z \rangle} \sin\theta  \nonumber \\
        &&+\ \frac{1}{2} A^2 \left.  \frac{\mathrm{d}^2P}{\mathrm{d}z^2} \right|_{z= \langle z \rangle} \sin^2\theta  \nonumber \\
        &&+\ \frac{1}{6} A^3 \left.  \frac{\mathrm{d}^3P}{\mathrm{d}z^3} \right|_{z= \langle z \rangle} \sin^3\theta  \nonumber \\
        &&+\ \frac{1}{24} A^4 \left.  \frac{\mathrm{d}^4P}{\mathrm{d}z^4} \right|_{z= \langle z \rangle} \sin^4\theta  \nonumber \\
        &&+\ \frac{1}{120} A^5 \left.  \frac{\mathrm{d}^5P}{\mathrm{d}z^5} \right|_{z= \langle z \rangle} \sin^5\theta  \nonumber \\
        &&+\ \frac{1}{720} A^6 \left.  \frac{\mathrm{d}^6P}{\mathrm{d}z^6} \right|_{z= \langle z \rangle} \sin^6\theta + \cdots,  
\end{eqnarray}
where $\theta \equiv 2\pi ft$.

As shown in the main text, we define $(-1)^n \left.  \frac{\mathrm{d}^nP}{\mathrm{d}z^n} \right|_{z= \langle z \rangle}$ as $P^{(n)}(\langle z \rangle)$.
This definition including coefficient $(-1)^n$ is compatible with the typical display style of SNOM approach curves $S_n(\langle z \rangle)$; when the scattering light increases exponentially as the tip approaches (i.e., $\mathrm{d}z < 0$), the corresponding $P^{(n)}(\langle z \rangle)$ represent exponential curves similar to $P(z)$, i.e., $P^{(n)}(\langle z \rangle) > 0$ at low $z$.

Using multiple-angle formulas, $\sin^m\theta$ can be expanded as follows;
\clearpage
\begin{eqnarray}
    P[z(t)] &=&  P(\langle z \rangle)  \nonumber \\
        &&- A P^{(1)}(\langle z \rangle) \sin\theta  \nonumber \\
        &&+\ \frac{1}{4} A^2 P^{(2)}(\langle z \rangle) \left( 1 - \cos 2\theta \right)  \nonumber \\
        &&-\ \frac{1}{24} A^3 P^{(3)}(\langle z \rangle) \left( 3 \sin\theta - \sin 3\theta \right)  \nonumber \\
        &&+\ \frac{1}{192} A^4 P^{(4)}(\langle z \rangle) \left( 3 - 4\cos 2\theta + \cos 4\theta \right)  \nonumber \\
        &&-\ \frac{1}{1920} A^5 P^{(5)}(\langle z \rangle) \left( 10 \sin\theta - 5\sin 3\theta + \sin 5\theta \right)  \nonumber \\
        &&+\ \frac{1}{23040} A^6 P^{(6)}(\langle z \rangle) \left( 10 - 15\cos 2\theta + 6 \cos 4\theta - \cos 6\theta \right) +\cdots.
        \label{eq_SI_Psincos}
\end{eqnarray}
\noindent
As described in Eq.~2 of the main text, the lock-in signal corresponds to 
\begin{eqnarray}
    S_n(\langle z \rangle) = \left\langle P[z(t)] \sin(n\theta + \phi_n) \right\rangle
    \label{eq_SI_Stimeavg}
\end{eqnarray}
\noindent
for the $n$-th harmonics.
Among the components of $\sin(m\theta)$ or $\cos(m\theta)$ in Eq.~\ref{eq_SI_Psincos}, only the component at $m=n$ gives a non-zero time-averaged value, as $\langle \sin^2 n\theta \rangle = \langle \cos^2 n\theta \rangle = \frac{1}{2}$.
For example, the first harmonic signal $S_1$ is maximized at $\phi_1 = \pi$ as 
\begin{eqnarray}
    S_1(\langle z \rangle) &=& \left\langle P[z(t)] (-\sin\theta) \right\rangle \nonumber \\
    &=& A P^{(1)}(\langle z \rangle)  \left\langle \sin^2\theta  \right\rangle \nonumber \\
    &&+\ \frac{1}{24} A^3 P^{(3)}(\langle z \rangle) \left\langle 3\sin^2\theta  \right\rangle \nonumber \\
    &&+\ \frac{1}{1920} A^5 P^{(5)}(\langle z \rangle) \left\langle 10 \sin^2\theta \right\rangle + ... \nonumber \\
    &=& \frac{1}{2} A P^{(1)}(\langle z \rangle) + \frac{3}{48} A^3 P^{(3)}(\langle z \rangle) + \frac{1}{384} A^5 P^{(5)}(\langle z \rangle) +\cdots.
    \label{eqS1}
\end{eqnarray}
\noindent
In a similar manner, each lock-in signal is maximized at $\phi_n = \frac{n+1}{2} \pi$ as
\begin{eqnarray}
    S_2(\langle z \rangle)  &=& \frac{1}{8} A^2 P^{(2)}(\langle z \rangle) + \frac{1}{96} A^4 P^{(4)}(\langle z \rangle) + \frac{1}{3072} A^6 P^{(6)}(\langle z \rangle) +\cdots, \label{eqS2} \\
    S_3(\langle z \rangle)  &=& \frac{1}{48} A^3 P^{(3)}(\langle z \rangle) + \frac{1}{768} A^5 P^{(5)}(\langle z \rangle) +\cdots, \label{eqS3} \\
    S_4(\langle z \rangle)  &=& \frac{1}{384} A^4 P^{(4)}(\langle z \rangle) + \frac{1}{7680} A^6 P^{(6)}(\langle z \rangle) +\cdots.  \label{eqS4} 
\end{eqnarray}
\noindent
Equation~3 in the main text corresponds to a simplified description of Equations~\ref{eqS1} to \ref{eqS4}.
Note that practically, depending on the lock-in amplifier, the experimental phase values $\phi_n$ can shift from the theoretical values.
Assuming that $P^{(5)}$, $P^{(6)}$, and higher order derivatives are negligibly small, the $n$-th derivative of the scattering laser power $P$ with respect to $\langle z \rangle$ is described using the lock-in signals $S_n$ as follows:
\begin{eqnarray}
    P^{(1)}(\langle z \rangle) &\approx& \frac{2}{A} \left[S_1(\langle z \rangle) - 3S_3(\langle z \rangle) \right],  \label{eqP1}\\
    P^{(2)}(\langle z \rangle) &\approx& \frac{8}{A^2} \left[S_2(\langle z \rangle) - 4S_4(\langle z \rangle) \right],  \label{eqP2}\\
    P^{(3)}(\langle z \rangle) &\approx& \frac{48}{A^3} S_3(\langle z \rangle),  \label{eqP3} \\
    P^{(4)}(\langle z \rangle) &\approx& \frac{384}{A^4} S_4(\langle z \rangle).  \label{eqP4}
\end{eqnarray}

Figures~\ref{figS3}A, D, G, and J show the approach curves of $S_1$ to $S_4$, respectively, recorded with several oscillation amplitudes $A = 0.1$--5.0 nm.
Using different $A$ provides different curve appearances, but the curves can be matched by the normalization. 
Figures~\ref{figS3}B, E, H, and K show the $P^{(1)}(\langle z \rangle)$ to $P^{(4)}(\langle z \rangle)$ curves calculated from Eqs.~\ref{eqP1} to \ref{eqP4}, respectively.
Each $n$-th derivative value is consistent with any $A$ used.
On the one hand, small $A$ causes severe noise at higher $n$ due to the coefficient of $1/A^n$ in the equations.
For example, in the third (fourth) derivative channel, the normalized curve(s) with $A = 0.1$ nm ($A = 0.5$ and 0.1 nm) is (are) too noisy to read the signal appearance like an exponential function [Figs.~\ref{figS3}H and J (K and L)].
On the other hand, with a too large $A$, higher order harmonics components is no longer negligible in $S_n$ signals.
For example, as shown in Eq.~\ref{eqS1}, the $S_1$ curve has the contribution of $A^3P^{(3)}(\langle z \rangle)$ in addition to the major factor of $AP^{(1)}(\langle z \rangle)$.
As a result, the raw approach curves with $A = 5.0$ nm (Figs.~\ref{figS3}A and D) appear to deviate substantially from the curves with the other smaller $A$. 
Figure~\ref{figS3}C (\ref{figS3}F) show the first (second) derivative curve reproduced by only considering $S_1$ ($S_2$) signals, i.e., assuming that $\mathrm{d}^nP/\mathrm{d}z^n \approx  n!(-2/A)^n S_n$.
In this rough approximation, the calculated curve with $A = 5.0$ nm does not match those with the other $A$ (Figs.~\ref{figS3}C and F).
In contrast, as described above, the more accurate approximation using Eqs.~\ref{eqP1} and \ref{eqP2} improves the matching (Figs.~\ref{figS3}B and E).
Note that the calculated $P^{(1)}$ curve with $A = 5.0$ nm still deviates slightly from the other curves, possibly due to the contribution of higher harmonics, i.e., $A^5P^{(5)}$ in Eq.~\ref{eqS1}.
Therefore, from the dataset, $A = 1.0$ nm is the optimal condition under which all the $S_1$ to $S_4$ channels well reproduce the $n$-th differential curves with a reasonable signal-to-noise ratio.
Note that as mentioned in the main text, the optimal amplitude value is expected to depend on the experimental setup used, such as the tip apex structure and the efficiency of the light collection.

\subsubsection*{Supplementary Text 2: ULA-SNOM mapping over a Ag atomic step}

The Ag(111) surface partially covered by ultrathin Si islands was first characterized by overview STM imaging without cantilever oscillation (Fig.~\ref{figS4}A).
The appearance is in good agreement with the previous STM study of amorphous Si films on Ag(111) \cite{solonenko2017comprehensive}.

In addition to the Si islands on an identical Ag terrace (Fig.~4 in the main text), we obtained ULA-SNOM maps including a Ag step. 
As schematically shown in Fig.~\ref{figS4}B, the scanned area has two Ag terraces separated by a single Ag-atomic step, both of which are partially covered by Si islands.
The bare Ag terraces (regions I and III), Si islands (II and IV), and the monoatomic Ag step (between II and III) are easily identified by the contrast in the STM image (Fig.~\ref{figS4}C).
The appearances of the four regions are useful to understand the contrast in each image. 
Regions I and II have the same Ag terrace height, whereas regions I and III have the same Ag surface but with a different height by one Ag atomic layer, and regions I and IV have the same atom-layer numbers with different atomic species (Ag and Si) in the topmost layer.
While the STM feedback ensures a constant tip--sample gap distance on the same material (e.g., regions I and III), the Ag-tip--Ag-surface gaps are slightly narrower by $\sim$50 pm than the Ag--Si gaps (orange dotted curves in Fig.~\ref{figS4}B). 

The images of $S_1$ and $S_2$ (Figs.~\ref{figS4}E and \ref{figS4}F) are not sensitive to the presence of the Si islands, as shown in Figs.~4C and 4D in the main text; however, the images show a contrast with regions I and II (III and IV) apparently darker (brighter), similar to the contrast-inverted surface topography.
The topography-synchronized contrast in $S_1$ and $S_2$ is probably caused by the adjustment of the tip--sample distances at the two terraces across the Ag step (Fig.~\ref{figS4}B) and scattering contributions from less localized fields.

At higher harmonics $S_3$ and $S_4$ (Figs.~\ref{figS4}G and \ref{figS4}H), the image contrast changes as the Si islands in regions II and IV now exhibit darker than the Ag terraces in regions I and III, which is consistent with the mapping results over an identical Ag terrace (Figs.~4E and 4F in the main text). 
The high resolution s-SNOM mapping enables us for optical surface analysis in the $\sim$1-nm scale.
For example, based on the contrast of the $S_4$ image, the bright spot with a diameter of $\sim$2 nm (marked by the dotted circles in Figs.~\ref{figS4}C and \ref{figS4}H) is not composed of Si, although it is located on an Si island, but can rather be assigned to an Ag cluster.

On the maps, we also observed the effect of an atomic-scale structural change in the tip on the images.
We observed an accidental tip change during the scan (black arrows in Figs.~\ref{figS4}C to \ref{figS4}H), as evidenced by a sudden jump of the simultaneously recorded STM and FM-AFM signals.
Because the same islands and the Ag step were observed similarly even after the change, this is attributed to a change in the atomic structure of the tip apex.
Notably, this minor tip change has a notable impact on the contrast in $S_3$ and $S_4$, but only a minor effect on $S_1$. 
This indicates that the atomic structure of the Ag tip apex  tunes the plasmonic field in the tip--sample gap, leading to the modification of the optical contrast in the s-SNOM maps.
Such atomic-level structures potentially forms plasmonic picocavities \cite{benz2016single,barbry2015atomistic}, which also contributed to the plasmonic field intensity and confinment in the tip--sample junction.
Revealing the existence of picocavities and utilizing them would further improve the spatial resolution of ULA-SNOM.



\begin{figure} 
	\centering
	\includegraphics[width=0.64\textwidth]{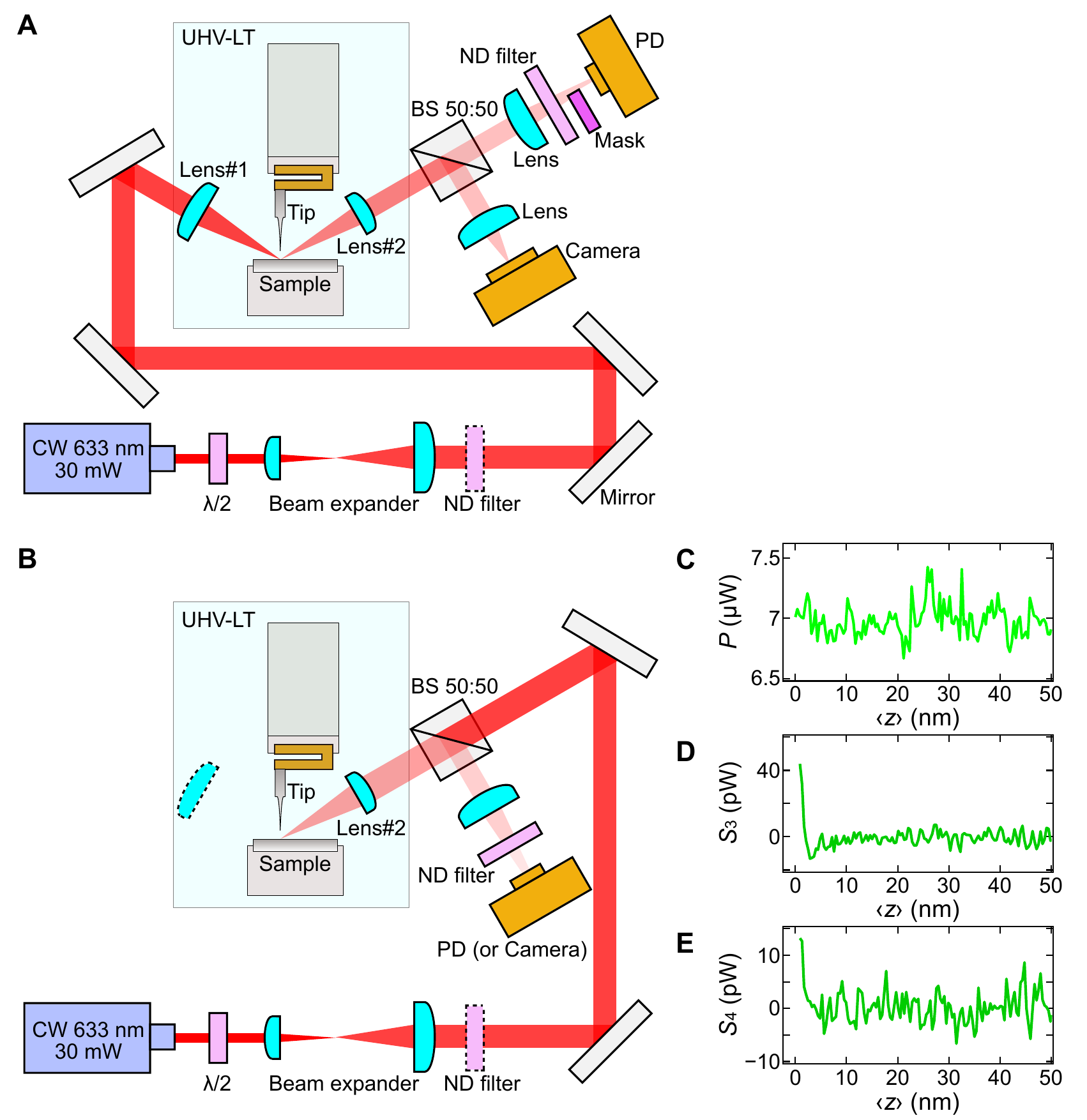} 

	\caption{\textbf{Two optical configurations for ULA-SNOM.}
	(\textbf{A}) Configuration we mainly used.
    Continuous-wave 633-nm laser output entered into the UHV chamber and was focused into the tip--sample junction via Lens \#1 (NA = 0.4) mounted in the STM/FM-AFM unit.
    The scattering light from the junction was collected by Lens \#2 (NA = 0.2) together with reflecting light and was directed outside the chamber to a PD through an neutral density (ND) filter and a mask. 
    To adjust the scattering light intensity to a PD-detectable range, the mask was used for blocking the strong light directly reflecting by the sample independent of the tip existence.
    (\textbf{B}) Another optical configuration of a back-scattering geometry.
    Lens \#2 was used both for incident light focusing and for scattering light collection.
    The scattering light was directed to the PD via a beam splitter (BS).
    (\textbf{C} to \textbf{E}) Approach curves of $P$, $S_3$, and $S_4$, respectively, recorded over a bare Ag(111) terrace with the configuration in (B).
    The high harmonic s-SNOM channels have signals in a very close tip--sample gap, which is in good agreement with the results with the configuration in (A).
    This verifies that both configurations are capable of ULA-SNOM measurements.
    Nevertheless, the s-SNOM signals in the latter configuration (B) are generally weaker than those in the former (A) because in the latter configuration, the incident laser intensity is reduced by passing through the BS.
    }
	\label{figS1} 
\end{figure}

\begin{figure} 
	\centering
	\includegraphics[width=0.75\textwidth]{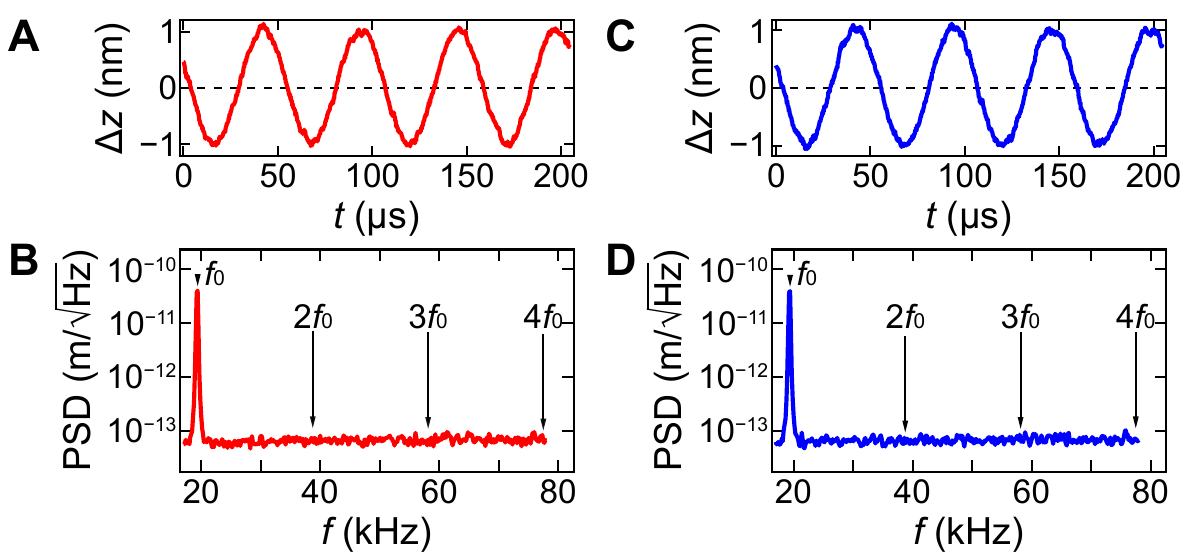} 

	\caption{\textbf{Monitoring the cantilever oscillation of the QTF sensor with an oscilloscope and spectrum analyzer.}
	(\textbf{A}) Time trace of the cantilever oscillation $\Delta z(t)$ and (\textbf{B}) its power spectral density (PSD).
    The Ag tip was far from more than 350 nm from the Ag(111) surface.
    The cantilever oscillation was excited by a piezo actuator with the resonance frequency, \textit{i.e.}, $f = f_0$, and the oscillation amplitude $A$ was set at 1.0 nm.
    (\textbf{C}) Time trace and (\textbf{D}) PSD after the tip was approached to the sample.
    The tip height was controlled by the STM feedback with sample bias $V_\mathrm{s}$ = 30 mV, time-averaged tunneling current $\langle I_\mathrm{t} \rangle$ = 0.10 nA with the cantilever oscillation, which is the same set-point as that for the s-SNOM maps shown in  Fig.~4 of the main text.
    Because the frequency shift $\Delta f$ at the tip height is much smaller than $f_0$ ($|\Delta f / f_0| < 0.04$\%), the position of the resonance peak in the spectrum appears unchanged from that in (B). 
    Both before and after the tip approach, the oscillation shows pure harmonicity without detectable higher-harmonics components ($2f_0$, $3f_0$, and $4f_0$), as indicated in (B) and (D).
    This verifies that the lock-in signals of the scattering light (Fig.~2) and their difference in the maps (Fig.~4) at low tip heights do not originate from anharmonic cantilever motion but the near-field.
    This result is in constant to anharmonic oscillation of a Si cantilever with tapping-mode AFM at low tip heights \cite{hillenbrand2000higher,esteban2009full}.
    We note that the detection of higher harmonic components of the cantilever oscillation with a QTF sensor was reported \cite{hembacher2004force}; however, the sensitive detection is owed to special properties of the sensor (the enhancement of the piezoelectric current output at high frequencies) \cite{giessibl2006higher,giessibl2019qplus}.
    The amplitudes of higher harmonics are intrinsically very small ($\sim$0.1 pm for $A$ = 1 nm) \cite{giessibl2006higher} due to the stiffness of the cantilever (a spring constant of 1800 N/m, two orders of magnitude larger than Si cantilevers).
    }
	\label{figSrev1} 
\end{figure}

\begin{figure} 
	\centering
	\includegraphics[width=0.6\textwidth]{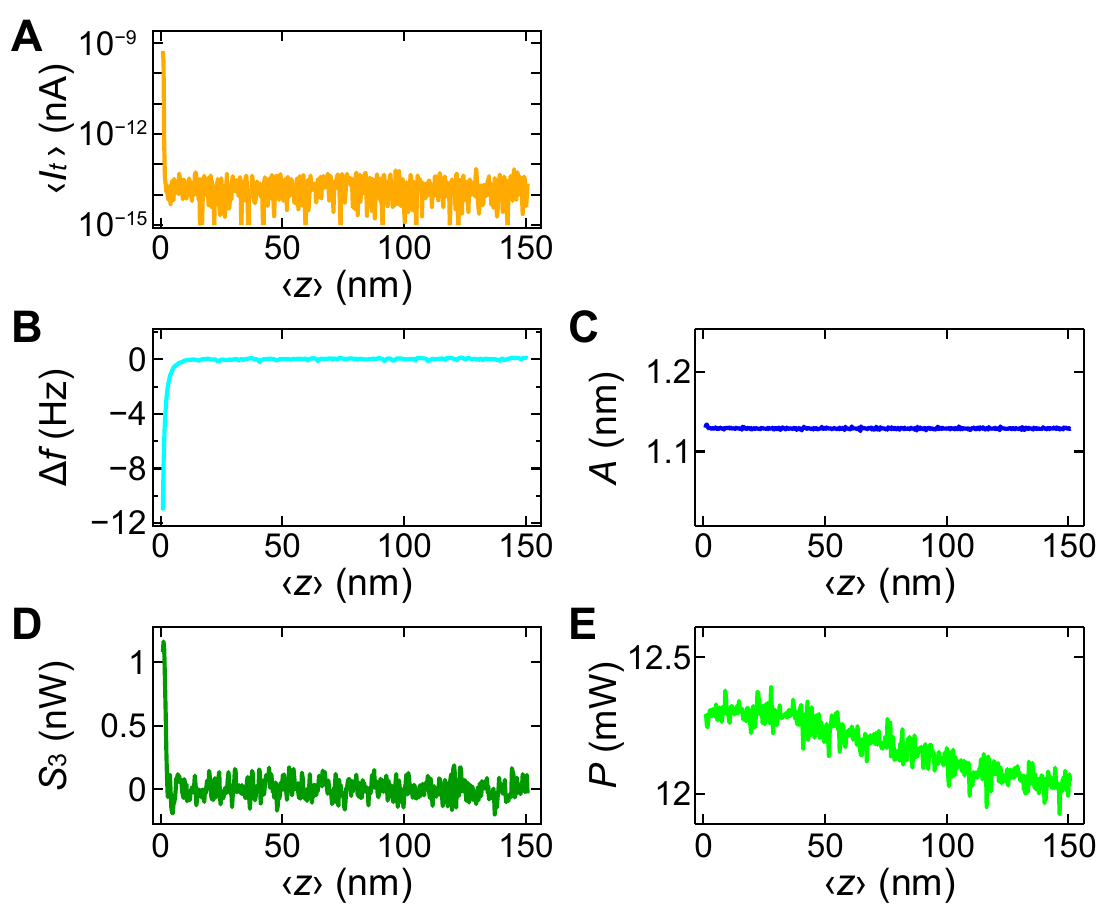} 

	\caption{\textbf{The same approach curves as those shown in Figs.~2B to 2D of the main text, but with the full tip-height range recorded.}
	(\textbf{A}) $\langle I_\mathrm{t} \rangle$, (\textbf{B}) $\Delta f$, (\textbf{C}) $A$, (\textbf{D}) $S_3$, and (\textbf{E}) $P$.
    }
	\label{figS2} 
\end{figure}

\begin{figure} 
	\centering
	\includegraphics[width=0.8\textwidth]{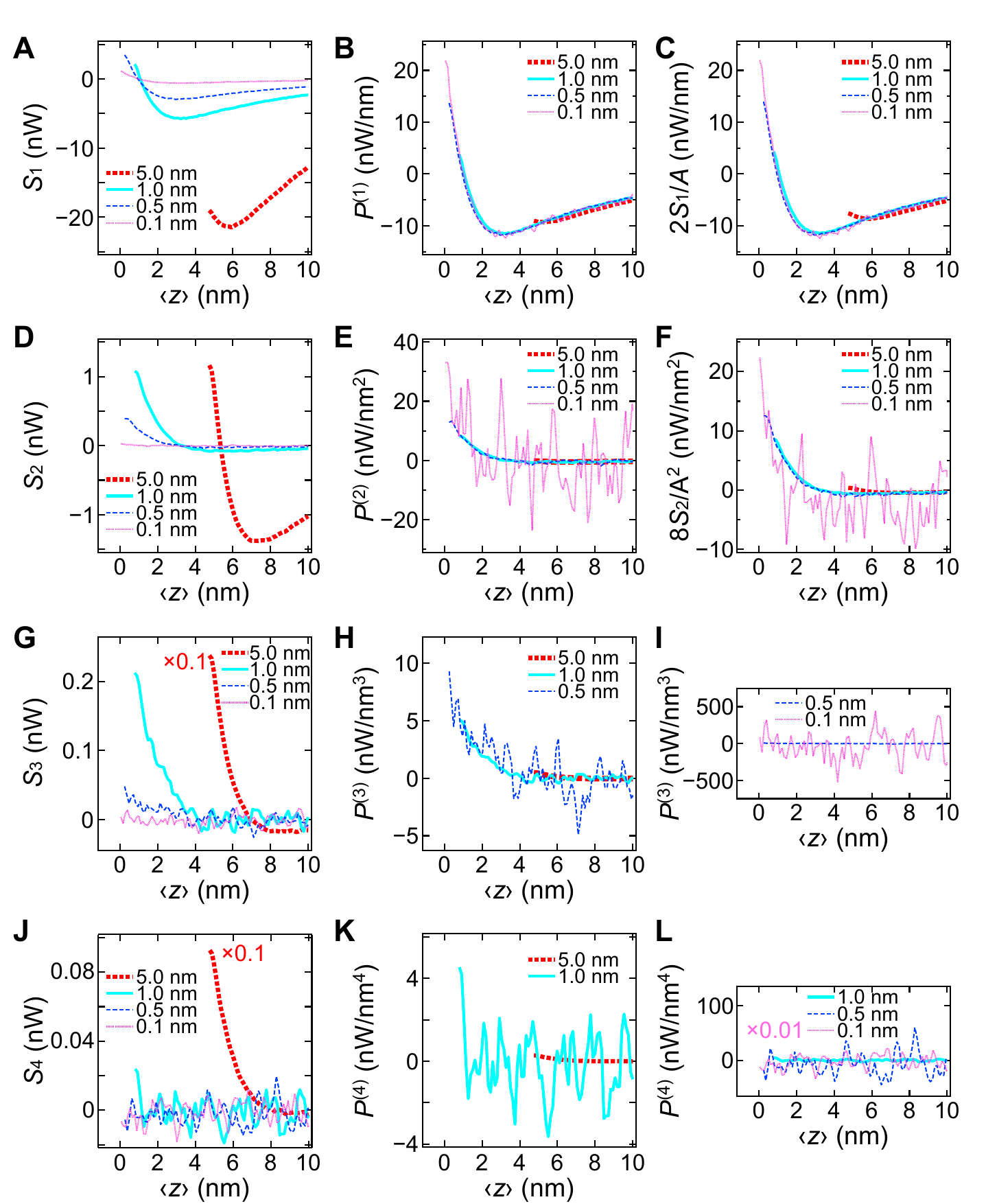} 

	\caption{\textbf{Approach curves of $S_n$ with different $A$ and their normalized curves.}
	(\textbf{A}) $S_1$, (\textbf{B}) $P^{(1)}$ calculated from $S_1$ and $S_3$, and (\textbf{C}) $P^{(1)}$ calculated only from $S_1$.
    (\textbf{D}) $S_2$, (\textbf{E}) $P^{(2)}$ calculated from $S_2$ and $S_4$, and (\textbf{F}) $P^{(2)}$ calculated only from $S_2$.
    (\textbf{G}) $S_3$, (\textbf{H}) $P^{(3)}$ calculated from  $S_3$ with larger $A$, and (\textbf{I}) $P^{(3)}$ calculated from  $S_3$ with lower $A$.
    (\textbf{J}) $S_4$, (\textbf{K}) $P^{(4)}$ calculated from $S_4$ with larger $A$, and (\textbf{L}) $P^{(4)}$ calculated from  $S_4$ with lower $A$.
    The plots in (G) to (I) are the same as those shown in Figs.~3E and 3F of the main text.
    }
	\label{figS3} 
\end{figure}

\begin{figure} 
	\centering
	\includegraphics[width=0.7\textwidth]{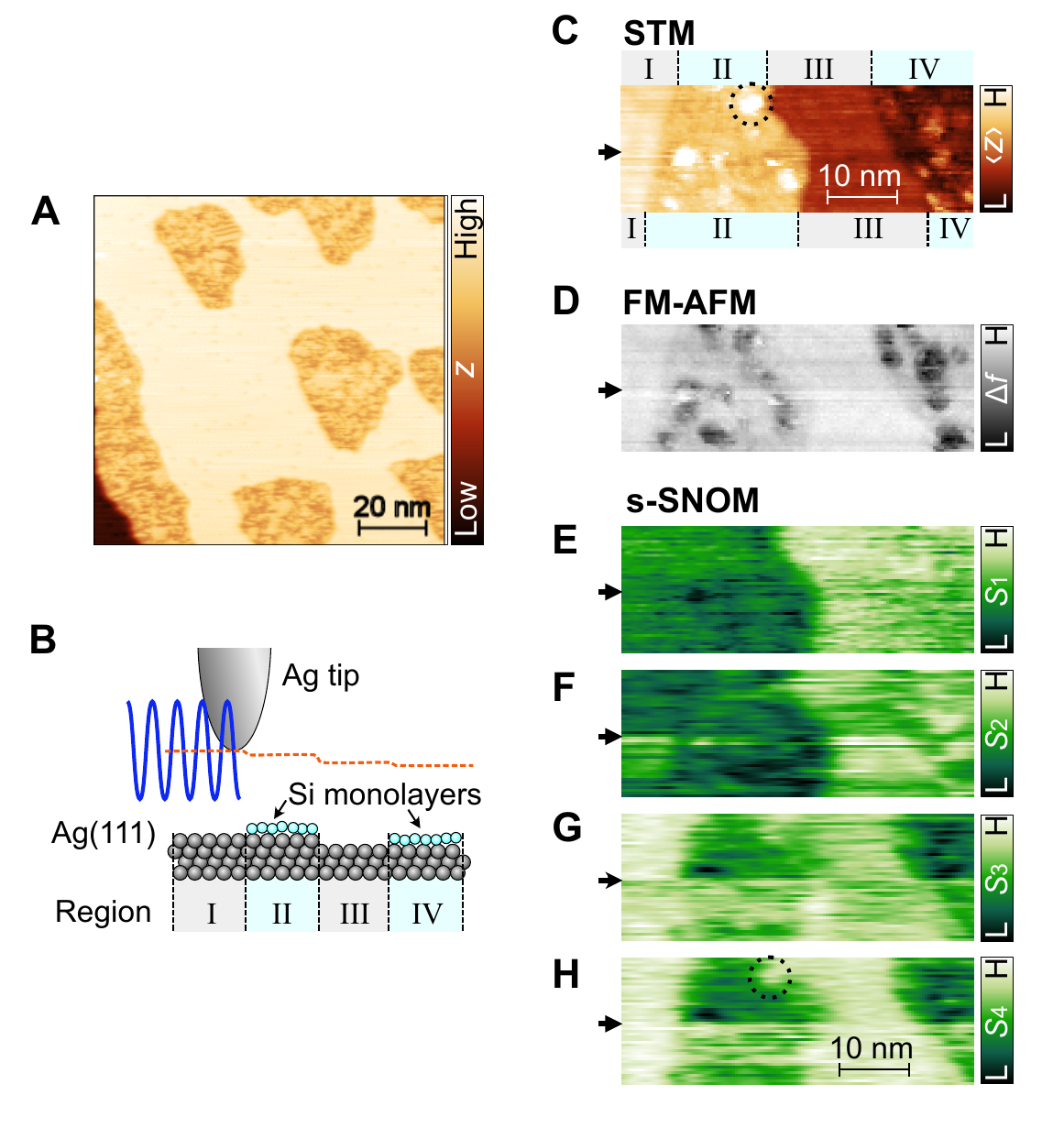} 

	\caption{\textbf{Simultaneously acquired STM, FM-AFM, and s-SNOM images of a different area from that for Figs.~4A to 4F in the main text.} 
    (\textbf{A}) Typical STM image of amorphous Si monolayer islands on Ag(111) obtained with the FIB-polished Ag tip without laser illumination (set-point: $V_\mathrm{s}$ = 1 V and $I_\mathrm{t}$ = 0.1 nA without cantilever oscillation).
	(\textbf{B}) Side-view scheme of the atomic structures of the sampling area.
    Regions I--IV correspond to the bare Ag on the upper terrace, Si island on the upper terrace, bare Ag on the lower terrace, and Si island on the lower terrace, respectively.
    The orange dotted line schematically indicates the tip-height trajectory.
    (\textbf{C}) STM topography, (\textbf{D}) FM-AFM $\Delta f$ map, and (\textbf{E} to \textbf{H}) s-SNOM $S_1$ to $S_4$ maps simultaneously obtained with laser illumination (STM set-point: $V_\mathrm{s}$ = 30 mV, $\langle I_\mathrm{t} \rangle$ = 0.10 nA, $A$ = 1.0 nm; $P_\mathrm{inc} = 6$ mW).
    The arrows indicate the position when the tip-apex structure accidentally changed.
    The black dotted circles in (C) and (H) mark a nanocluster in region II.
    }
	\label{figS4} 
\end{figure}

\begin{figure} 
	\centering
	\includegraphics[width=0.7\textwidth]{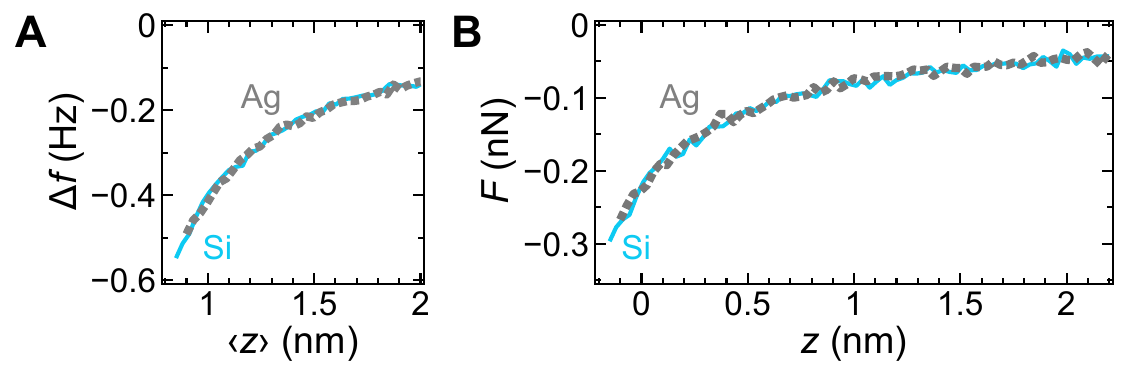} 

	\caption{\textbf{Frequency shift and force curves simultaneously recorded with s-SNOM approach curves}
	(\textbf{A}) $\Delta f(\langle z \rangle)$ recorded over Ag (dotted gray curve) and Si (solid cyan curve).
    Each curve was simultaneously obtained with the s-SNOM approach curve $S_4(\langle z \rangle)$ shown in Fig.~4H of the main text.
     The origin of $\langle z \rangle$ for both curves is defined by the STM set-point over the Ag terrace, as well as that in Fig.~4H.
    (\textbf{B}) $F(z)$ curves converted from the curves in (A).
    While the $S_4$ intensity on Si differs from that on Ag at small $\langle z \rangle$ (Fig.~4H), $\Delta f$ and $F$ recording over the two locations are comparable at any tip heights.
    }
	\label{figSrev2} 
\end{figure}



\end{document}